\documentclass[%
preprint,
 amsmath,amssymb,
 aps,
]{revtex4-2}

\usepackage[normalem]{ulem}  % for \sout
\usepackage{xcolor}  

\usepackage{graphicx}                   % for insert file eps
\usepackage{hyperref}                   % for hyper refrences
\usepackage{amsmath,amssymb}            % for insert extra math
\usepackage{multirow}	
\usepackage{xcolor}	
\usepackage{amsmath,color}

\graphicspath{{s/}}					% the path of your fig folder

\DeclareMathAlphabet{\mathpzc}{OT1}{pzc}{m}{it}
\begin{document}

	\author{Abbas Ali Saberi}
	\email{saberi@pks.mpg.de}
	\affiliation{Department of Physics, University of Tehran, P. O. Box 14395-547, Tehran, Iran}
	\affiliation{Max Planck Institute for the Physics of Complex Systems, 01187 Dresden, Germany}
 
     \author{Ugur Tirnakli}
     \email{ugur.tirnakli@ieu.edu.tr}
	 \affiliation{Department of Physics, Faculty of Arts and Sciences, Izmir University of Economics, 35330 Izmir, Turkey}
	
	\author{Constantino Tsallis}
 \email{tsallis@cbpf.br}
	\affiliation{Centro Brasileiro de Pesquisas Fisicas and National Institute of Science and Technology of Complex Systems, Rua Xavier Sigaud 150, Rio de Janeiro, Brazil}
\affiliation{Santa Fe Institute, 1399 Hyde Park Road, New Mexico 87501, USA}
\affiliation{Complexity Science Hub Vienna, Metternichgasse 8, 1030 Vienna, Austria}

	\title{Central Limit Behavior at the Edge of Chaos in the $z$-Logistic Map %Family
    }
	%\email{author's email address}
	
	\date{\today}
	
	\begin{abstract}
    We focus on the Feigenbaum–Coullet–Tresser point of the dissipative one-dimensional $z$-logistic map $x_{t+1} = 1 - a |x_t|^z \;\;(z \ge 1)$. We show that sums of iterates converge to $q$-Gaussian distributions $P_q(y)=P_q(0)\,\exp_q(-\beta_q\, y^2)= P_q(0)\, [1+(q-1)\beta_q\, y^2]^{1/(1-q)}\;(q \ge 1\,; \beta_q > 0)$, which optimize the nonadditive entropic functional $S_q$  under simple constraints. We derive a closed-form prediction for the entropic index,
$q(z)=1+2/(z+1)$, and validate it numerically via data collapse for typical $z$ values. The formula captures how the limiting law depends on the nonlinearity order and implies finite variance for $z>2$ and divergent variance for $1\le z\le2$. These results extend edge-of-chaos central-limit behavior beyond the standard ($z=2$) case and provide a simple predictive law for unimodal maps with varying maximum order.

    %We focus on the dissipative one-dimensional $z$-logistic map $x_{t+1} = 1 - a |x_t|^z \;\;(z \ge 1; \,0 \le a \le 2; \,\, x_t \in [-1,1]; \, t=0,1,2,\dots)$, which is topologically equivalent to the standard logistic map ($z=2$). At its $z$-generalized Feigenbaum-Coullet-Tresser point, where its successive bifurcations accumulate, the Lyapunov exponent vanishes (weak chaos). By summing a large number of iterates starting at different values of the initial condition $x_0$, we numerically exhibit a generalized Central Limit Theorem behavior, with the attractor (in the space of probability distributions) emerging to be the $q$-Gaussian distribution $P_q(u)=P_q(0)e_q^{-\beta u^2}= P_q(0) [1+(q-1)\beta u^2]^{1/(1-q)}\;(q \ge 1)$ which, under simple constraints, optimizes the nonadditive entropic functional $S_q=\frac{1-\sum_{i=1}^W p_i^q}{q-1} \;\left(S_1=-\sum_{i=1}^W p_i \ln p_i\right)$. By heuristically assuming a divergence condition, we obtain and numerically verify $q(z)=1+2/(z+1)$, which implies a finite (diverging) variance for $z>2$ ($1 \le z \le 2$).
	\end{abstract}
	\maketitle

%%%%%%%%%%%%%%%%%%%%%%%%%
\section{Introduction}
\begin{figure}
 \centering
 \includegraphics[width=8.15cm]{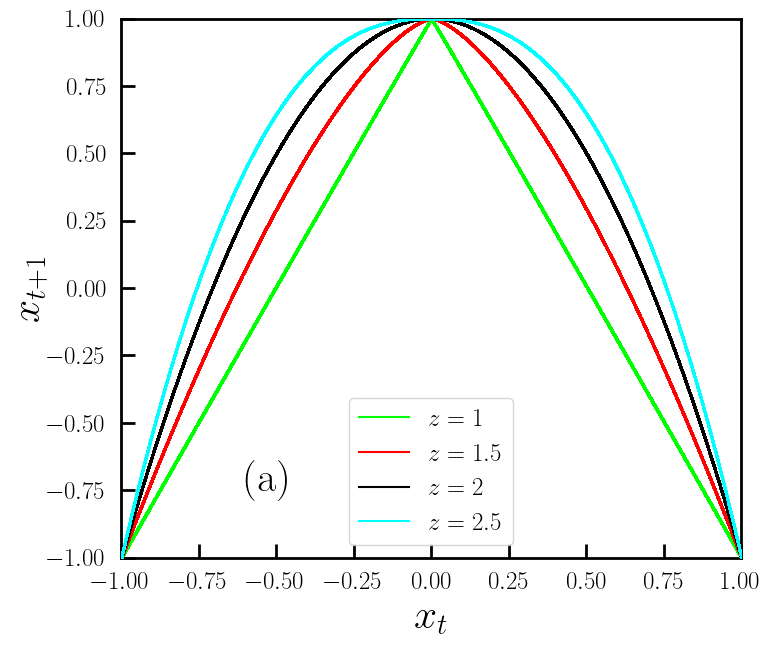}
 \includegraphics[width=8.15cm]{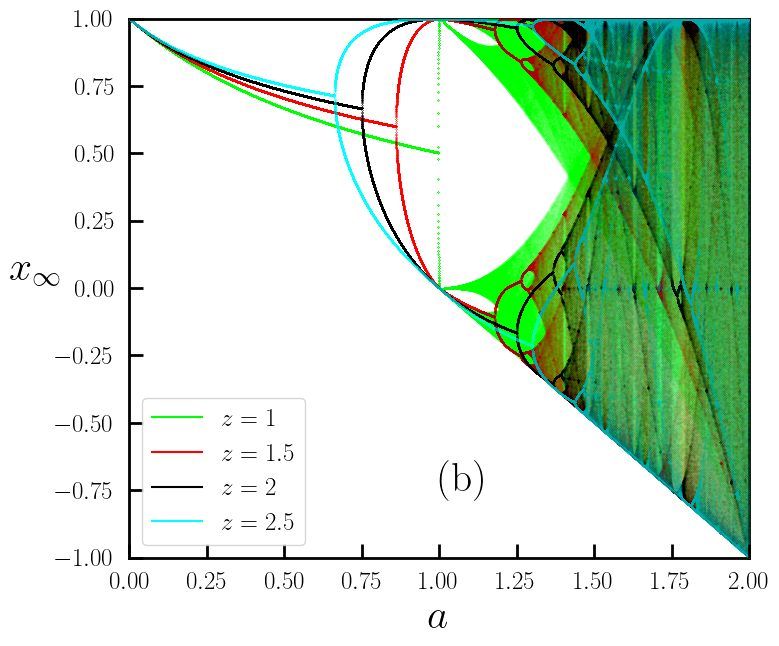}
 \caption{(a)~Phase portrait and (b)~bifurcation diagram of the $z$-logistic map
for some representative values of $z$.} 
 \label{xvsx}
\end{figure}

%%%%%%%%%%%%%%%%%%%%%%%%%
In the study of dynamical systems theory, one of the most important and widely used workhorses is, no doubt, the logistic map. For discrete cases, it is a fundamental model commonly used to study population dynamics, chaos, and bifurcation phenomena. Mathematicians mostly like to use the definition given as
\begin{equation}
    X_{t+1} = r \, X_t (1 - X_t) \;\;\;\;\;\;(0 \leq r \le 4; \, X_t \in [0,1]; \, t=0,1,2,\dots)
    \label{eq:logistic_map1}
\end{equation}
where \( r \) is a parameter that governs the behavior of the map (frequently called control parameter), and \( X_t \) is the state variable at iteration \( t \) \cite{Strogatz,Hilborn}. It has long been known that the logistic map exhibits a variety of dynamical behavior, ranging from fixed points to chaotic regimes with periodic windows as the parameter \( r \) is varied. 
On the other hand, most physicists prefer to use the definition in the form
\begin{equation}
    x_{t+1} = 1 - a |x_t|^2 \;\;\;\;\;\;(0 \le a \le 2; \,\, x_t \in [-1,1]; \, t=0,1,2,\dots)\,,
    \label{eq:logisticmap2}
\end{equation}
where $a$ and $x_t$ are the new map parameter and the state variable, respectively. It is easy to show that the definitions (\ref{eq:logistic_map1}) and (\ref{eq:logisticmap2}) are topologically conjugated and can be transformed into each other using a conjugation function \cite{Beckbook}. 
Various types of generalization for the standard logistic map can be found in the literature \cite{Hamada2024,Lawnik2017,Moon2008,HuMao1982,Hu1982,Hauser1984,Weele1987,Briggs1991}. Here, we study the generalized version based on the definition in Eq.~\eqref{eq:logisticmap2}, introducing an additional parameter \( z \) which allows us to control the degree of non-linearity of the system at its maximal point. Therefore, the \textit{generalized \( z \)-logistic map} is defined as follows:
\begin{equation}
    x_{t+1} = 1 - a |x_t|^z \;\;\;(z \ge 1; 0 \le a \le 2; \,\, x_t \in [-1,1]; \, t=0,1,2,\dots) \,,
    \label{eq:z_logistic_map}
\end{equation}
where \( z \) controls the power to which the state variable \( x_t \) is raised, introducing a wider range of topologically isomorphic dynamic behaviors, though metrically different from the classical logistic map, as can be seen from Fig.~\ref{xvsx}a. As \( z \) increases, the map becomes more chaotic in the sense that fixing \(a\) in the chaotic regime (e.g. \(a=2\)) and increasing \(z\) strengthen the chaos metrically \cite{baladi2000positive}. 
The parameter \( a \) governs the overall structure of the map. Similarly to the standard logistic map ($z=2$), varying \( a \) triggers transitions of the system from stable fixed points to periodic orbits and eventually chaotic behavior as depicted in Fig.~\ref{xvsx}b. From this figure and from an earlier work \cite{Lyra1997}, one can also easily see that the chaos threshold point $a_c(z)$ (defined as the critical point where the system enters into the chaotic region via period-doublings of successive bifurcations) tends to 1 as $z\to 1$, while it comes closer to 2 as $z\to\infty$, suppressing the chaotic region of the map.
Some values of $a_c$ are given in Table~\ref{table1} for representative values of $z$.

We provide a closed-form prediction for the nonadditive index that governs the central-limit attractor of sums of iterates of the $z$-logistic map at the Feigenbaum–Coullet–Tresser point. Specifically, we predict
\begin{equation}
    q(z)=1+\frac{2}{z+1} \;\;\;(z \ge 1)\, .
\label{qzz}    
\end{equation} This relation is obtained from a tail–moment argument tied to the order $z$ of the map’s maximum. The index is not fitted; accordingly, the limiting $q$-Gaussian has a fixed shape (no free shape parameter), and only an overall scale is estimated from data. Using Huberman–Rudnick scaling to coordinate the approach to criticality with the number of summands, we validate this prediction across several $z$ values by data collapse of rescaled sums onto the corresponding $q$-Gaussians. Importantly, the $q(z)$ introduced here characterizes the statistics of sum distributions at the edge of chaos and should not be confused  with indices such as $q_{\mathrm{sen}}$ that quantify sensitivity to initial conditions; such indices need not to coincide in this setting.

\begin{table}[h]
    \centering
   \caption{$a_c$, $q$, $\bar{\beta}_q$ and $\delta_F(z)$ values for various $z$ of the $z$-logistic map.}
\label{table1}
    \begin{tabular}{llllll}
      \hline
      \hline
\multicolumn{1}{l|}{$z$} & \multicolumn{1}{l|}{$a_c$} & \multicolumn{1}{l|}{$q$} & \multicolumn{1}{l|}{$\bar{\beta}_q$} & \multicolumn{1}{l}{$\delta_F(z)$}   \\ \hline      \hline
      \multicolumn{1}{l|}{$1$} & \multicolumn{1}{l|}{1} & \multicolumn{1}{l|}{2} & \multicolumn{1}{l|}{$\pi^2$} & \multicolumn{1}{l}{$-$}  \\ \hline
      \multicolumn{1}{l|}{$1.75$} & \multicolumn{1}{l|}{1.35506075662336...} & \multicolumn{1}{l|}{1.727...} & \multicolumn{1}{l|}{6.489...} & \multicolumn{1}{l}{4.2571...}  \\ \hline
        \multicolumn{1}{l|}{$1.85$} & \multicolumn{1}{l|}{1.37474875895489...} & \multicolumn{1}{l|}{1.702...} & \multicolumn{1}{l|}{6.278...} & \multicolumn{1}{l}{4.4261...}  \\ \hline
      \multicolumn{1}{l|}{$2$} & \multicolumn{1}{l|}{1.40115518909205...} & \multicolumn{1}{l|}{1.667...} & \multicolumn{1}{l|}{6.002...} & \multicolumn{1}{l}{4.6992...} \\ \hline
      \multicolumn{1}{l|}{$2.15$} & \multicolumn{1}{l|}{1.42456107409069...} & \multicolumn{1}{l|}{ 1.635...} & \multicolumn{1}{l|}{5.768...} & \multicolumn{1}{l}{4.9017...}  \\ \hline
      \multicolumn{1}{l|}{$2.25$} & \multicolumn{1}{l|}{1.43880058172258...} & \multicolumn{1}{l|}{1.615...} & \multicolumn{1}{l|}{5.630...} & \multicolumn{1}{l}{ 5.0518...}  \\ \hline
     \multicolumn{1}{l|}{$\infty$} & \multicolumn{1}{l|}{2} & \multicolumn{1}{l|}{1} & \multicolumn{1}{l|}{$\pi$} & \multicolumn{1}{l}{$-$}  \\ \hline\hline
    \end{tabular}
\end{table}

%%%%%%%%%%%%%%%%%%%%%%%%%
\section{Central Limit Behavior}
%%%%%%%%%%%%%%%%%%%%%%%%%
Now we are in a position to discuss the Central Limit (CL) behavior of this system. As is well known, the classical CL theorem applies to sums of independent and identically distributed (i.i.d.) random variables. In any kind of real or model systems with such random variables, as the number of variables increases, the distribution of their sum converges to a Gaussian distribution. More precisely, one can write

\begin{equation}
    y = \frac{1}{\sqrt{T}} \sum_{i=1}^{T}(x_i-\left<x\right>)
    \label{sumy}
\end{equation}
where $T$ is the number of summand, $x$ is the random variable of the system considered and $\left<...\right>$ defines an average over a large number $T$ of iterations and a large number of randomly chosen initial values $x_1$, which can be numerically calculated as

\begin{equation}
    \left<x\right> = \frac{1}{n_{ini}}\frac{1}{T} \sum_{j=1}^{n_{ini}}
    \sum_{i=1}^{T}x_i^{(j)} \,.
    \label{ave}
\end{equation}

In our simulations, we discard the first $T_{\mathrm{trans}}=2^{15}$ iterates as transients and average over $n_{\mathrm{ini}}$ uniformly sampled initial conditions in $[-1,1]$; results are robust to alternative choices (e.g., deterministic grids). In the case of the $z$-logistic map, any chaotic point in the phase space, for example, for the ($z=2,\,a=2$) case, $y$ values would fulfill the condition for the random variables to be i.i.d. distributed since in this case the map is (semi)conjugated to a Bernoulli shift and strongly mixing. Therefore, one should expect to have a Gaussian probability distribution as $T\to\infty$, namely

\begin{equation}
    P(y) = \frac{1}{\sqrt{2\pi\sigma^2}} \exp{\left(-\frac{y^2}{2\sigma^2}\right)} \,.
    \label{Py}
\end{equation}
This has already been studied in \cite{Tirnakli2007}, and Gaussian behavior is observed. In fact, this behavior has also been achieved for other parameter values in the chaotic region. One can argue that the standard CL theorem will be valid if the Lyapunov exponent is positive. 
On the other hand, when we concentrate on the chaos threshold point where the standard Lyapunov exponent vanishes, the map is not sufficiently strongly mixing and, due to ergodicity breaking, the standard CL theorem is not valid anymore. In order to study the CL behavior of such systems, the standard CL theorem has recently been generalized \cite{CLT1,CLT2,CLT3,CLT4,UmarovTsallis2022}. 
Consistent with generalized central-limit behavior for strongly correlated dynamics, the following q-Gaussian attractors are expected/observed under appropriate scaling,

\begin{equation}
P_q(y)=\frac{\sqrt{\beta_q}}{C_q} \exp_q\left(-\beta_q y^2\right) ,
\label{qGaussian}
\end{equation}
where $\beta_q$ is a parameter that characterizes the width of the distribution and here $\exp_q(u)$ is known as $q$-exponential of the form $\left[1+(1-q)u\right]^{\frac{1}{1-q}} \;\;(q\ge 1)$. The normalization factor $C_q$ can be calculated from \cite{prato-tsallis-1999,Tsallisbook} and is given by
\begin{equation}
	\label{eq:Cq}
	C_q=\left\{\begin{array}{lc}
		\displaystyle \sqrt{ \pi }&q=1 \,,\\
		\displaystyle\frac{\Gamma\left(\frac{3-q}{2q-2}\right)} {\Gamma \left(\frac{1}{q-1}\right)} 
		\sqrt{\frac{\pi}{q-1}} &1<q<3 \, .
	\end{array} \right.
\end{equation}
Let us remind at this point that distributions (\ref{qGaussian}) optimize, under simple constraints, the nonadditive entropic functional \cite{Tsallis1988}
\begin{equation}
S_q[P(y)] =\frac{1-\int dy \, [P(y)]^q}{q-1} \;\;\;(q \in \mathbb{R}) \,,
\end{equation}
which, in the $q\to 1$ limit, recovers the classical Boltzmann-Gibbs (additive) functional
\begin{equation}
S_1[P(y)] =S_{BG}[P(y)]\equiv -\int dy P(y) \ln P(y) \,.
\end{equation}

If the prefactor in front of the $q$-exponential in Eq.~(\ref{qGaussian}) is defined as $P_q(0)$, after some algebra, one can rewrite the same equation as 

\begin{equation}
	\label{qGaussian2}
	\frac{P_q(y)}{P_q(0)} =  
    \left[1-(1-q)\bar{\beta}_q(y P_q(0))^2\right]^{\frac{1}{1-q}} \, .
\end{equation}

The subtle difficulty in analyzing the system at the edge of chaos (i.e., at $a=a_c$) is that taking $T\to\infty$ alone is not sufficient for the system to achieve its limit distribution.
The other ingredient must be to localize the critical point $a_c$ with infinite precision. More precisely, for a full description of the shape of the distribution function on the attractor, a simultaneous limit is needed, as the precision of the $a_c$ value and the number of iterates tend to infinity. However, in numerical experiments, neither the precision of $a_c$ nor the values of $T$ can reach infinity. Therefore, numerically, one needs to focus on the situation as a kind of interplay between the precision of $a_c$ and the number of iterates. In \cite{Afsar2013}, it has been shown that a successful interplay can be achieved using the Huberman-Rudnick scaling law \cite{HR1980}. Basically, the idea is to find $(a,T)$ tuples from 
\begin{equation}
	\label{eq:HR}
	2^{-n} = |a-a_c(z)|^{\ln2 / \ln\delta_F(z)} 
\end{equation}
where $\delta_F(z)$ is the $z$-generalized Feigenbaum constant (calculated numerical values are in Table~\ref{table1}) and $n=1,2,...,\infty$ which will be used to find the value $T$ to be used ($T=2^{2n}$) for a particular value of $a$ in the vicinity of $a_c(z)$ (see \cite{Afsar2013} for more details). All $(a,T)$ tuples used in this work are given in Table~\ref{tablez}. Here, we set all values of the parameter $a$ so that the precision of the corresponding $2n$ value would be of the order of $0.05$. 
The CL behavior of the standard logistic map ($z=2$) has already been numerically analyzed in this way in \cite{Tirnakli2009,Afsar2013} and the value of $q$ is found to be close to 1.65.

\begin{table}[h]
\begin{center}
\label{tablez}
\caption{Parameter values used in this work. The values of $(a,T)$ tuples are obtained from the scaling Eq.~(\ref{eq:HR}). } 
%\vspace{0.6cm}
\begin{tabular}{|l|l|l|c|l|}
  \hline
\hline 
 $z$  &   $a$   &  $2n$   & $T=2^{2n}$  \\ \hline \hline
    &  $1.3550696971$&  16.05    & $2^{16}$        \\ \cline{2-4}
\cline{2-4}        
  1.75  & $1.3550628567$ &  18.05    & $2^{18}$    \\ \cline{2-3}
\cline{2-4}
     & $1.3550612499$ &  20.05    & $2^{20}$              \\ \hline
\hline   
    &  $1.3747552996$ &  16.05    & $2^{16}$        \\ \cline{2-3}
\cline{2-4}         
  1.85  & $1.3747502366$ &  18.05    & $2^{18}$    \\ \cline{2-3}
\cline{2-4}
     & $1.3747490928$ &  20.05    & $2^{20}$ 
          \\ \hline
\hline   
    &  $1.4011592349$ &  16.05    & $2^{16}$        \\ \cline{2-3}
\cline{2-4}         
  2  & $1.4011560500$ &  18.05    & $2^{18}$    \\ \cline{2-3}
\cline{2-4}
     & $1.4011553723$ &  20.05    & $2^{20}$ 
          \\ \hline
\hline   
    &  $1.4245639575$ &  16.05    & $2^{16}$        \\ \cline{2-3}
\cline{2-4}         
  2.15  & $1.4245616623$ &  18.05    & $2^{18}$    \\ \cline{2-3}
\cline{2-4}
     & $1.4245611940$ &  20.05    & $2^{20}$ 
          \\ \hline
\hline   
    &  $1.4388028452$ &  16.05    & $2^{16}$      \\ \cline{2-3}
\cline{2-4}
  2.25   & $1.4388010297$ &  18.05    & $2^{18}$   \\ \cline{2-3}
\cline{2-4}
     & $1.4388006704$ &  20.05    & $2^{20}$    \\ \hline
  
\end{tabular}
\end{center}
\end{table}

%%%%%%%%%%%%%%%%%%%%%%%%%
\section{Relation between $q$ and $z$}
%%%%%%%%%%%%%%%%%%%%%%%%%
In order to relate $q$ and $z$ values of the $z$-logistic map in Eq.~\ref{eq:z_logistic_map}, let us assume that the moment $\langle |x|^z \rangle$ of a $q$-Gaussian is proportional to $\int_0^\infty dx \,|x|^z \, \exp_q(-x^2)$, which diverges (logarithmically) for 

\begin{equation}
\frac{2}{q-1} -z =1 \;\;\;(z\ge 1) \,.
\label{eq:scaling}
\end{equation}
From this condition, we obtain $q=5/3=1.666...$ for $z=2$. This value obtained from the scaling relation is in fact very close to what has been found earlier in \cite{Tirnakli2009,Afsar2013} for the standard logistic map. Moreover, the case $z=1$ (a piecewise-linear unimodal map) yields $q=2$ from our prediction; this is consistent with increasingly heavy tails at lower $z$ and it has already been known analytically in the literature \cite{Lai2000} that this is the genuine distribution for this case. Therefore, we have the second point corroborating the scaling given in Eq.~(\ref{eq:scaling}). 
It might well be that this scaling applies to {\it all values of} $z$. Therefore, the CL behavior of the probability distribution at the $z$-generalized Feigenbaum-Coullet-Tresser point would be expected to approach a $q$-Gaussian with the index $q$ {\it a priori} known from the equation given as indicated in Eq. (\ref{qzz}).
%\begin{equation}
%q(z) = 1+\frac{2}{z+1} \;\;\;(z\ge 1)  \,. 
%\label{qz}
%\end{equation}
Consequently, $q$ monotonically decreases from 2 to 1 when $z$ increases from 1 to infinity.
In what follows, we provide numerical support to this conjecture. 
In the generalized CL theorem framework \cite{CLT1,CLT2}, sums of variables whose effective variance diverges converge, after appropriate rescaling, to the $q$‑Gaussian in Eq.~(\ref{qGaussian2}).
For $q>1$, $q$‑Gaussian has a power‑law tail.  Setting $u=(q-1)\beta_q\,y^2\gg1$ in Eq.~\eqref{qGaussian} yields
\begin{equation}
P_q(y)\sim u^{\,1/(1-q)} \sim y^{-\frac{2}{q-1}}.
\label{eq:tail}
\end{equation}
Thus, the asymptotic exponent is $2/(q-1)$. The absolute $m$‑th moment is
\begin{equation}
\langle |y|^m\rangle
\sim 2\int_{Y_0}^\infty y^m\,y^{-\frac{2}{q-1}}\,dy
= 2\int_{Y_0}^\infty y^{\,m-\frac{2}{q-1}}\,dy.
\end{equation}
Convergence at the upper limit requires
\begin{equation}
m-\frac{2}{q-1}<-1,
\end{equation}
while the threshold of (logarithmic) divergence is
\begin{equation}
m-\frac{2}{q-1}=-1
\quad\Longrightarrow\quad
q-1=\frac{2}{m+1}.
\label{eq:m_threshold}
\end{equation}

In the standard CL theorem, for i.i.d. linear variables, one takes $m=2$ (the variance).  For the map \eqref{eq:z_logistic_map}, however, the iteration involves $|x|^z$, so the \emph{natural} moment whose divergence governs the attractor class is the $z$‑th moment: $m\;\longrightarrow\;z$.
Imposing Eq.~\eqref{eq:m_threshold} with $m=z$ directly yields
\begin{equation}
q-1=\frac{2}{z+1} \,,
%\quad\Longrightarrow\quad
%q(z)=1+\frac{2}{z+1}.
\end{equation}
which implies, in turn, Eq. (\ref{qzz}).

The final step before trying to corroborate the proposed scaling relation with numerical experiments remains to determine the $\bar{\beta}_q$ values, since we already know the values of $q$ for any $z$. In fact, it is easy to find the $\bar{\beta}_q$ values once we notice that, in order to obtain Eq.~\eqref{qGaussian2} from Eq.~\eqref{qGaussian}, we take $P_q(0)=\sqrt{\beta_q}/C_q$, which immediately makes $\beta_q y^2 = C_q^2(yP_q(0))^2$ and from there it is clear that
\begin{equation}
\bar{\beta}_q=C_q^2\,.
\label{eq:betaq}
\end{equation}
Some representative values of $(q,\bar{\beta}_q)$ are given in Table~\ref{table1}.

%%%%%%%%%%%%%%%%%%%%%%%%%
\section{Numerical validation of $q(z)$}
%%%%%%%%%%%%%%%%%%%%%%%%%
Now we are ready to present our numerical results in order to validate our scaling relation \ref{qzz} and the corresponding $\bar{\beta}_q$ value from Eq.~(\ref{eq:betaq}). It is worth noting here that, for a particular $z$ value, $q$ and $\bar{\beta}_q$ are not fitting parameters; they are known a priori when we start simulations.
Therefore, the only fitting parameter in these simulations is $P_q(0)$. In order to numerically approach this value, let us start from Eq.~(\ref{qGaussian2}) and assume that $(q-1)\bar{\beta}_q P_q(0)^2 y^2 \gg 1$, which allows us to write
\begin{equation}
P_q(0) = (q-1)^{1/(q-3)} \bar{\beta}_q^{1/(q-3)} \lim_{y\to\infty} P_q(y)^{(q-1)/(q-3)} y^{2/(q-3)} \, .
\label{eq:Pq0}
\end{equation}
In the numerics, for a better estimate, we take an average over the values obtained for large $y$ values before the sharp drop is attained. 

\begin{figure}
 \centering
 \includegraphics[width=\columnwidth]{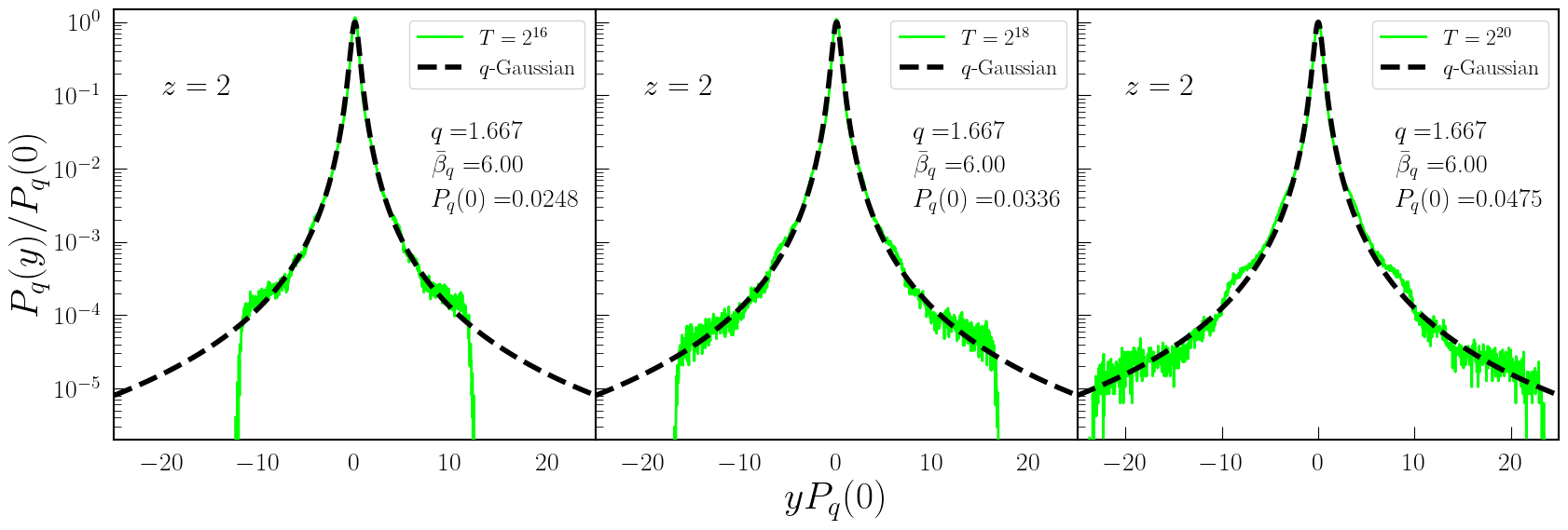}
 \caption{Normalized probability distribution given by Eq.~(\ref{qGaussian2}) with increasing $T$ for $z=2$. For the histogram, we always use 1000 boxes to have equally sized bins for the entire time series from minimum to maximum $y$ values. Notice that here the only fitting parameter for the $q$-Gaussian (black dashed line) is $P_q(0)$.
 } 
 \label{z2N}
\end{figure}

\begin{figure}
 \centering
 \includegraphics[width=\columnwidth]{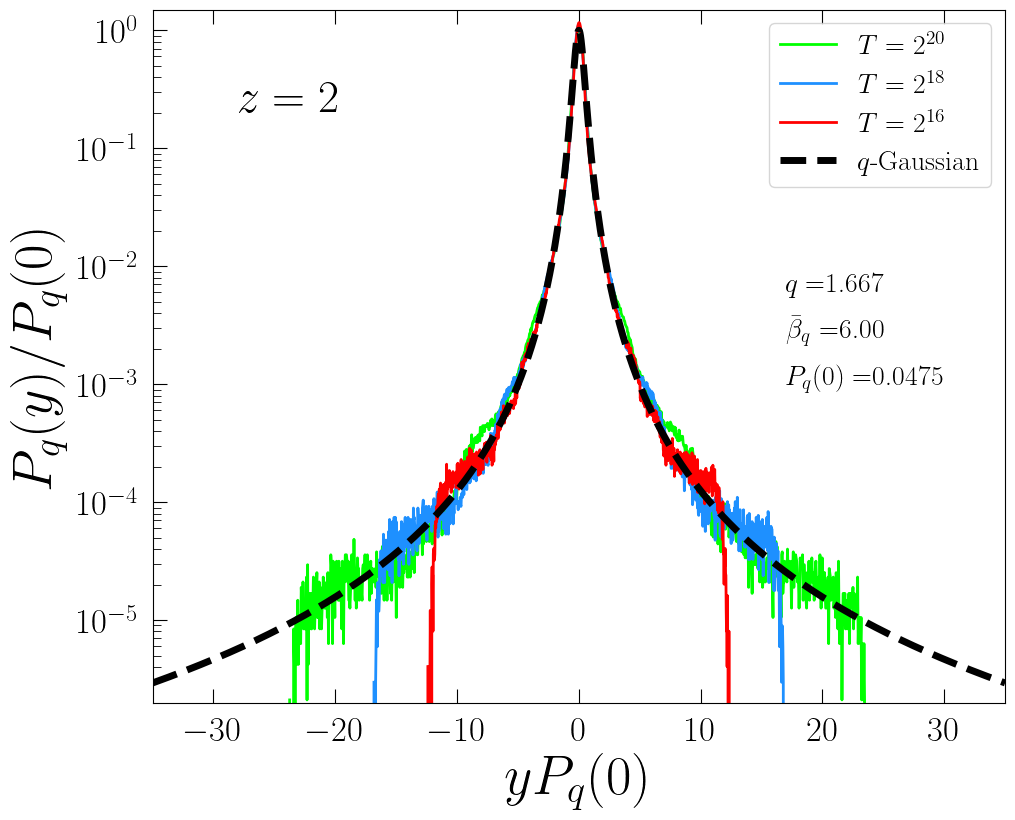}
 \caption{Data collapse of Fig.\ref{z2N}. It is evident that, as $(a,T)$ tuple simultaneously approaches $a\to a_c$ and $T\to\infty$, the simulation results become developing in the tails more and more. 
 } 
 \label{z2}
\end{figure}

Naturally, as a first case, we reconsider the standard logistic map ($z=2$). It is evident from Fig.~\ref{z2N} that, as the values of $T$ increase (and the related value $a$ given in Table~\ref{tablez} is used), the probability distribution develops in the tails of the $q$-Gaussian whose $(q,\bar{\beta}_q)$ pair comes a priori directly from our Eq.~(\ref{qzz}) and Eq.~(\ref{eq:betaq}). In Fig.~\ref{z2}, we also plot these three cases together so that one can better see the development in the tails of the appropriate $q$-Gaussian curve. 
Up to now, we have three $z$ values that corroborate the proposed scaling relation \ref{qzz}: (i)~$z\to 1$ gives $q=2$ (Cauchy distribution), (ii)~$z\to\infty$ leads $q=1$ (Gaussian distribution), and (iii)~$z=2$ indicates $q=5/3$.
As $z\to\infty$ and $a$ is fixed, $|x|^z\to 0$ for $|x|<1$, so $x_{t+1}\to 1-a$ (a nearly constant map). In our parametrization, the predicted $q(z)\to 1$ reflects thinning tails (Gaussian limit).

\begin{figure}
 \centering
 \includegraphics[width=\columnwidth]{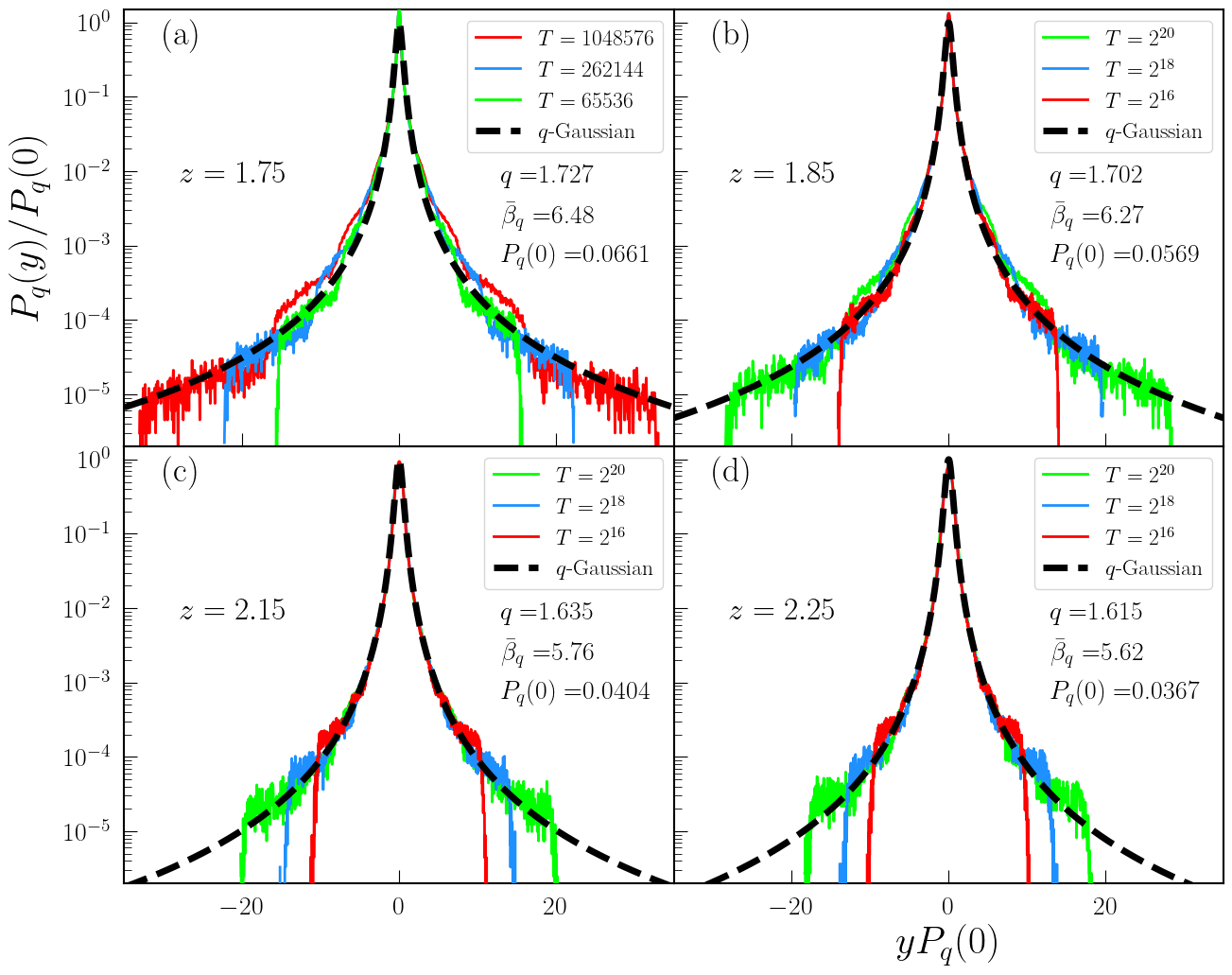}
 \caption{Data collapse of three different $(a,T)$ tuples for the cases $z=1.75, 1.85, 2.15, 2.25$. The computational cost corresponding to values of $z$ very distant from $z=2$ exceeds our present capacities.
 } 
 \label{fig4z}
\end{figure}

\begin{figure}
 \centering
 \includegraphics[width=\columnwidth]{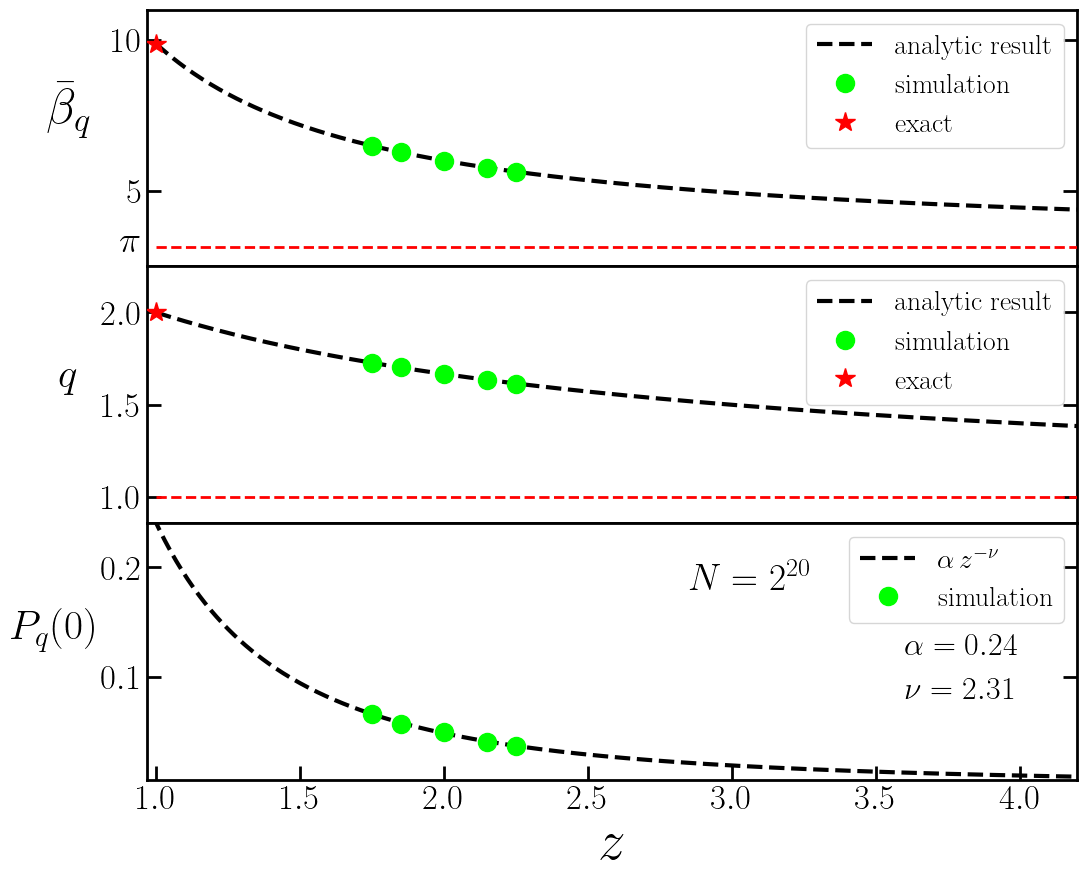}
 \caption{Black curves are analytic and green dots are from numerics. We have that $\lim_{z\to\infty} (\bar{\beta}_q,q)=(\pi,1)$ and $\lim_{z\to 1} (\bar{\beta}_q,q)=(\pi^2,2)$. The tentative heuristic relation $P_q(0) = \alpha z^{-\nu}$ has been obtained from the fitting of these 5 green numerically obtained points. Notice that, in contrast with $P_q(0)$, $q$ and $\bar{\beta_q}$ are not fitting parameters but are instead given by Eq.~(\ref{qzz}) and Eq.~(\ref{eq:betaq}) respectively. 
 } 
 \label{final}
\end{figure}

Now, let us numerically check four more $z$ values (two smaller, two larger than 2) to better corroborate the scaling relation. As the $z$ values differ from 2 along both directions, it becomes numerically harder to simulate. This is the reason why we choose $z$ values not too distant from 2. In all obtained cases, we use the quadruple precision instead of double precision and most probably, as $z$ values are too far away from 2, even higher precisions will be needed. For the chosen $z$ values ($z=1.75, 1.85, 2.15, 2.25$), the results are given in Fig.~\ref{fig4z}. For all cases, we use the $(q,\bar{\beta}_q)$ tuples obtained from Eq.~(\ref{qzz}) and Eq.~(\ref{eq:betaq}), which allows us to see a strong corroboration between the expected $q$-Gaussian and the numerically obtained histogram. In order to better visualize this corroboration, in Fig.~(\ref{final}), we plot the $z$ dependence of $\bar{\beta}_q$, $q$ and $P_q(0)$. For $q$ and $\bar{\beta}_q$, the analytical results from Eq.~(\ref{qzz}) and Eq.~(\ref{eq:betaq}) are depicted as dashed black lines. The simulation results are shown as green dots, while the exact results for the $z\to1$ case are denoted by red stars. It is also evident that, as $z\to\infty$, expected values ($(\bar{\beta}_q,q)=(\pi,1)$) are approached. Finally, for a tentative heuristic relation of $P_q(0)$ with respect to $z$, we suggest $P_q(0)=\alpha z^{-\nu}$ with $\alpha=0.24$ and $\nu=2.31$.

\section{Conclusion}
We examined the central-limit behavior for sums of successive iterates of the one-dimensional $z$-logistic map at the Feigenbaum–Coullet–Tresser accumulation point, where the Lyapunov exponent vanishes (weak chaos) and strong mixing breaks down. Guided by a tail–moment divergence criterion tied to the order $z$ of the map maximum, we proposed a closed-form prediction for the index of the limiting law,
$q(z)=1+2/(z+1)$ [Eq.~(\ref{qzz})]. Together with $\bar{\beta}_q=C_q^2$ [Eq.~(\ref{eq:betaq})], this fixes the $q$-Gaussian attractor up to an overall scale. Using Huberman–Rudnick scaling to couple the distance to criticality with the number of summands, we obtained data collapse for typical $z$ values around $2$, providing numerical support for the prediction without tuning the shape. The result organizes how the attractor evolves with $z$ and implies a finite variance for $z>2$ and a divergent one for $1\le z\le2$.
\\The $q(z)$ determined here characterizes the statistics of \emph{sums} (the central-limit attractor) and should not be confused with $q_{\mathrm{sen}}$ or other members of the $q$-triplet that quantify sensitivity or relaxation; such indices do not coincide in this setting. Our analysis consolidates and extends edge-of-chaos central-limit behavior beyond the standard ($z=2$) case to a family of unimodal maps with varying nonlinearity.

As a meaningful consequence of the present result we can mention that the relation conjectured for the solar wind observations \cite{BurlagaVinas2004,TsallisGellMannSato2005,Tsallisbook}, namely
\begin{equation}
q_{sen}=1-\frac{q_{stat} -1}{2q_{stat}-3} \,,
\label{qtriplet}
\end{equation}
does {\it not} apply to the present system. Indeed, for $z=2$, we have $q\equiv q_{stat}=5/3$ and $q_{sen}=0.2444877\dots$, which violate Eq. (\ref{qtriplet}). In other words, the algebra governing the edge of chaos of the $z$-logistic maps is {\it not} the same that appears to govern the $q$-triplet associated to the solar wind \cite{BurlagaVinas2004,TsallisGellMannSato2005,GazeauTsallis2019}.

Future work includes enlarging the $z$-range and precision, quantitative goodness-of-fit and model-selection tests against competing heavy-tailed laws, extension to other periodic windows, and exploring whether a renormalization-group argument can make the $m=z$ moment criterion rigorous. These directions would further clarify universality classes of weakly chaotic central-limit behavior.

\begin{acknowledgments} 
The numerical calculations reported in this paper were partially performed at TUBITAK ULAKBIM, High Performance and Grid Computing Center (TRUBA resources). U.T. is a member of the Science Academy, Bilim Akademisi, Turkey and supported by the Izmir University of Economics Research Projects Fund under Grant No. BAP-2024-07. 
C.T. is partially supported by CNPq and Faperj (Brazilian agencies). 
\end{acknowledgments}

%	\bibliography{refs} % bibliography

\end{document}